\documentclass[12pt,preprint,showpacs,superscriptaddress,pre]{revtex4-1}
\usepackage{amsmath}
\usepackage{amssymb}

\usepackage[T1]{fontenc}

\usepackage{mathptmx}  

\usepackage[dvips]{graphicx}
\usepackage[svgnames]{xcolor}
\usepackage[colorlinks=true,citecolor=blue,linkcolor=LimeGreen,citebordercolor=white,linkbordercolor=white]{hyperref}
\usepackage{enumerate}

\def\eq#1{{equation (\ref{#1})}}

\begin{document}

\title{Far tails of the density of states in amorphous organic semiconductors}

\author{S.V. Novikov}
\email{novikov@elchem.ac.ru}
\affiliation{A.N. Frumkin Institute of
Physical Chemistry and Electrochemistry, Leninsky prosp. 31,
119071 Moscow, Russia}
\affiliation{National Research University Higher School of Economics, Myasnitskaya Ulitsa 20, Moscow 101000, Russia}


\begin{abstract}
Far tails of the density of state (DOS) are calculated for the simple models of organic amorphous material, the model of dipolar glass and model of quadrupolar glass. It was found that in both models far tails are non-Gaussian. In the dipolar glass model the DOS is symmetric around zero energy, while for the model of quadrupolar glass the DOS is generally asymmetric and its asymmetry is directly related to the particular geometry of quadrupoles. Far tails of the DOS are relevant for the quasi-equilibrium transport of the charge carriers at low temperature. Asymmetry  of DOS in quadrupolar glasses means a principal inequivalence of the random energy landscape for the transport of electrons and holes. Possible effect of the non-Gaussian shape of the far tails of the DOS on the temperature dependence of carrier drift mobility is discussed.

\end{abstract}

\maketitle

\section{Introduction}

One of the fundamental characteristics defining charge transport properties of any disordered material is a density of states (DOS) $P(U)$, i.e. a  distribution density of random energy  $U$ of a charge carrier hopping in the bulk of the material. In some situations the very functional form of $P(U)$ could be used for a quick estimation of the relevant transport properties of the material. For example, if the DOS decays fast for $U\rightarrow -\infty$, then the non-dispersive quasi-equilibrium transport regime eventually takes place for $t\rightarrow \infty$ and the stationary density of occupied states $P(U)\exp\left(-U/kT\right)$ is developed (we assume here the Boltzmann statistics). Let the maximum of this distribution is located at $U_{eq}$. In quasi-equilibrium regime the limiting step for the charge transport is a carrier escape from deep states where carrier hops up in energy to perform a  corresponding transition in space. It is reasonable to assume that the final energy of such hop is located somewhere near the maximum $U_0$ of the DOS. Hence, a rough estimation of the temperature dependence of the carrier drift mobility $\mu(T)$ for low electric field should be
\begin{equation}\label{mu}
\mu\propto \exp\left(-\frac{U_0-U_{eq}}{kT}\right)
\end{equation}
with $U_0-U_{eq}$  serving as an effective activation energy. For the Gaussian DOS
\begin{equation}\label{mu}
P(U)=\frac{1}{\left(2\pi\sigma^2\right)}\exp\left(-\frac{U^2}{2\sigma^2}\right)
\end{equation}
typical for amorphous organic semiconductors  \cite{Bassler:15,Novikov:2532,Vandewal:125204,Tal:256405}, $U_{eq}=-\sigma^2/kT$ and $\mu\propto \exp\left[-\sigma^2/(kT)^2\right]$. This simple estimation gives an exact leading asymptotics for the low field mobility temperature dependence in 1D case \cite{Dunlap:542} and differs from the corresponding asymptotics in higher dimensions by the numeric factor $\simeq 1$ in the exponent \cite{Bassler:15,Deem:911,Novikov:4472}. More refined approach invoking the conception of so-called transport energy provides even better description of the dependence $\mu(T)$ \cite{Baranovskii:2699,Nenashev:93201}, at least for the materials with spatially non-correlated random energy landscape.

Experimental data on the mobility temperature dependence suggest that in organic materials $\sigma\simeq 0.1$ eV \cite{Bassler:15} and, hence, at the room temperature $\sigma/kT\simeq 4-5$. For some materials charge transport has been observed in time-of-flight experiments even for $\sigma/kT\simeq 6-7$ \cite{Borsenberger:3066}, though in that case it is highly dispersive. This means that typically the maximum of the occupied DOS, relevant for the quasi-equilibrium transport, is located at the tail of the initial DOS.

There is a strong evidence that in organic materials a major contribution to the total DOS has the electrostatic origin: it is produced by the interaction of charge carrier with randomly located and oriented static dipoles and quadrupoles. Estimation of the electrostatic $\sigma$ taking into account typical dipole and qudrupole moments and concentration of polar molecules provides values in reasonable agreement with experimental data \cite{Novikov:4472}. Models of dipolar glass (DG) and quadrupolar glass (QG) have been suggested for description of random energy landscape in amorphous organic materials. In these models it is assumed that there is no correlation between orientations of dipoles or quadrupoles. The models naturally produce highly spatially correlated random energy landscape necessary for realization of the specific Poole-Frenkel mobility field dependence $\ln\mu\propto E^{1/2}$ \cite{Novikov:14573,Dunlap:542,Novikov:954}.

\begin{figure}[htbp]
\includegraphics[width=2.5in]{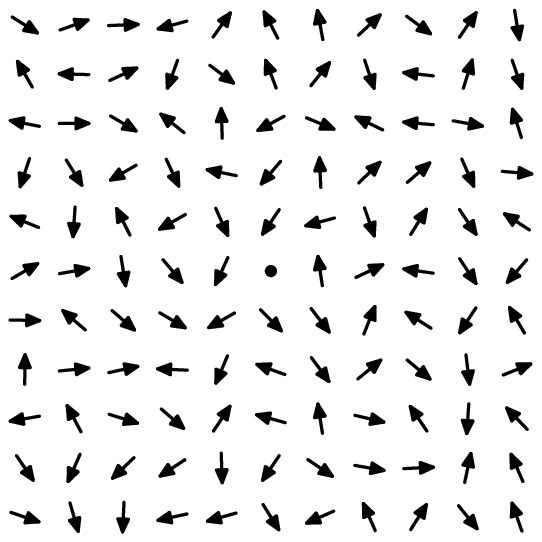}
\begin{center}
a
\end{center}
\vskip20pt
\includegraphics[width=2.5in]{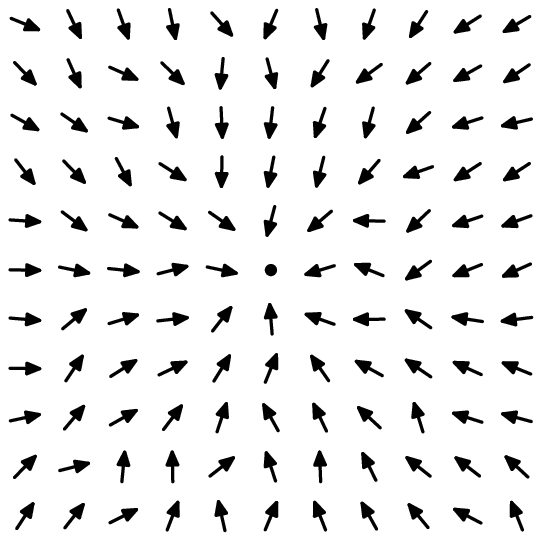}
\begin{center}
b
\end{center}
\caption{a) Typical distribution of dipoles that generates random energy $U$ in  the main body of the DOS (random energy is measured at the point in the center of the picture). b) Large stochastic cluster of dipoles relevant for the far tail of the DOS.} \label{2_typical_dist}
\end{figure}

Until now an accurate calculation of the behavior of the far tails of the DOS in DG and QG models has not been carried out. For DG and QG models the central peak of DOS has a Gaussian shape if concentration of dipoles and quadrupoles is not too low \cite{Dieckmann:8136,Novikov:877e,Novikov:954}. This shape is essentially guaranteed by the Central Limit Theorem which is not valid for the far tails of the distribution. Typical distribution of dipoles relevant for the main body of DOS  and for the far tail are qualitatively different (see Fig \ref{2_typical_dist} (a) and (b), correspondingly). In this paper we are going to carry out a direct calculation of the functional form of the tails of DOS in DG and QG models and estimate a possible effect of the tail shape on charge transport.

\section{Density of states in the DG model}

Let us consider the simple cubic lattice with sites occupied by randomly oriented dipoles having dipole moment $p$. We assume that there is no orientational correlation between dipoles, their orientations are uniformly distributed, and fraction of sites occupied by dipoles is $c$. Then the distribution density of the random energy $U$ is
\begin{eqnarray}\label{distribution}
P(U)=\left<\delta\left(U-\sum\limits_n U_n\right)\right> =\frac{1}{2\pi}\int\limits^\infty_{-\infty}dy\left<\exp\left[iy\left(U-\sum\limits_n U_n\right)\right]\right>=\\
=\frac{1}{4\pi}\int\limits^\infty_{-\infty}dy \exp(iyU)\prod_n\left[c\int\limits_{-1}^1 d(\cos\vartheta_n)
\exp(-iyU_n)+1-c\right],
\end{eqnarray}
where angular brackets mean an average over positions and orientations of dipoles, index $n$ runs over all lattice sites apart from the initial reference site and $U_n$ is the contribution of the point dipole at the site $n$ to the total energy $U$ of the charge carrier at the reference site
\begin{equation}\label{phi_n}
U_n=-\frac{e\vec{p}_n \vec{r}_n}{\varepsilon
r^3_n}=-\frac{ep}{\varepsilon r^2_n}\cos\vartheta_n,
\end{equation}
here $\varepsilon$ is a dielectric constant and $a$ is a lattice constant  \cite{Novikov:877e}. After integration over  $\vartheta_n$ in  \eq{distribution} we obtain \begin{equation}\label{distribution2}
P(U)=\frac{1}{2\pi}\int^\infty_{-\infty}dy
\exp\left[iyU+S(y)\right],\hskip10pt S(y)=\sum_n\ln\left(c\frac{\sin z_{n}}{z_{n}}+1-c\right),
\hskip10pt z_{n}=\frac{epy}{\varepsilon r^2_n}.
\end{equation}
Let us calculate the integral (\ref{distribution2}) using the saddle point method. The structure of the exponent in the integral (\ref{distribution2}) indicates that the saddle point is located at $y_s=i\omega$ where $\omega$ is a real number. We will see that for far tail $U\rightarrow \infty$ $\omega\rightarrow\infty$ too. We obtain a rough estimation of $S(\omega)$ replacing the summation in \eq{distribution2} with the integration
\begin{equation}\label{Sdip}
S(\omega)\simeq 2\pi z_a^{3/2}\int\limits_0^{z_a}\frac{dz}{z^{5/2}} \ln\left(c\frac{\sinh z}{z}+1-c\right),
\hskip10pt z_a=\omega U_d = \frac{ep\omega}{\varepsilon a^2}.
\end{equation}
For $\omega\rightarrow\infty$
\begin{equation}\label{Sdipasy}
S(\omega)=2\pi z_a^{3/2} \left[A_d -\int\limits_{z_a}^{\infty} \frac{dz}{z^{5/2}}\ln\left(c\frac{\sinh z}{z}+1-c\right)\right]\approx
2\pi z_a^{3/2} A_d-4\pi z_a,
\end{equation}
here we keep the leading correction only and
\begin{equation}\label{Ad}
A_d=\int\limits_0^{\infty} \frac{dz}{z^{5/2}}
\ln\left(c\frac{\sinh z}{z}+1-c\right).
\end{equation}
We find the position of the saddle point from the equation (note that $\omega > 0$ is appropriate for $U > 0$)
\begin{equation}\label{Spdip}
-U+\frac{dS}{d\omega}=-U+3\pi A_d U_d z_a^{1/2}-4\pi U_d=0,\hskip10pt z_a=\left(\frac{U+4\pi U_d}{3\pi A_d U_d}\right)^2
\end{equation}
and the saddle point approximation for the integral (\ref{distribution2}) is
\begin{equation}\label{Sp-app}
P(U)\approx \frac{1}{\left(2\pi \frac{d^2 S}{d\omega^2}\right)^{1/2}}\exp\left[-\omega U+S(\omega)\right].
\end{equation}
Substituting the solution of \eq{Spdip} in \eq{Sp-app}, we obtain the far tail asymptotics
\begin{equation}\label{Spdipas}
P(U)\approx \left(\frac{|U|+4\pi U_d}{9\pi^3 A_d^2 U^3_d}\right)^{1/2}
\exp\left[-\frac{\left(|U|+4\pi U_d\right)^3}{27\pi^2 A_d^2 U^3_d}\right],\hskip10pt |U|\rightarrow\infty,
\end{equation}
in this form the asymptotics is valid for both positive and negative $U$. We see that far tails are non-Gaussian but the DOS is still symmetric around $U=0$.

Saddle point approximation is valid if the correction to the quadratic term in the expansion of $S(\omega)$ around the maximum is negligible at the scale $\delta\omega\simeq \left(\frac{d^2 S}{d\omega^2}\right)^{-1/2}$. Estimating the correction using the next term of the expansion, we obtain the necessary condition for the validity of the approximation as $\left|\frac{d^3 S}{d\omega^3}\right|\left(\frac{d^2 S}{d\omega^2}\right)^{-3/2} \propto z_a^{-3/4} \ll 1$. This inequality is valid if $|U|/U_d \gg 1$. In fact, all other corrections are negligible, too.

\section{Density of states in the QG model}

In the QG model we replace dipoles with quadrupoles. A point quadrupole generates the electrostatic potential
\begin{equation}\label{Qpot}
\varphi(\vec{r})=\frac{\sum\limits_{i,j} Q_{ij}x_i x_j}{2\varepsilon r^5}.
\end{equation}
where $\textbf{Q}$ is a symmetric traceless tensor $\sum\limits_i Q_{ii}=0$. It may be transformed to the diagonal form $\textbf{Q}=\textrm{diag}(-Q(1+\alpha)/2,-Q(1-\alpha)/2,Q)$ having two scalar parameters $Q$ and $\alpha$. Any quadrupole could be considered as a linear combination of axial and planar quadrupoles $\textbf{Q}=\textbf{Q}_a-\alpha \textbf{Q}_p$/2, where $\textbf{Q}_a=\textrm{diag}(-Q/2,-Q/2,Q)$ and $\textbf{Q}_p=\textrm{diag}(Q,-Q,0)$. Let us consider two separate cases, i.e. the cases of pure axial (AQ) and planar (PQ) quadrupoles.

Note  that there is a very distinct difference between density of states in AQ and PQ glasses. For PQ the inversion $Q\rightarrow -Q$ that transforms the positive carrier energy to negative one and vice versa is equivalent to the spatial rotation of the quadrupole. All spatial configurations of the quadrupole have the equal weight, hence, the DOS is symmetric around $U=0$. For AQ the inversion is not equivalent to any possible rotation and the DOS is not symmetric. Substantial asymmetry is developing only for far tails of the DOS because for typical high concentration of quadrupoles the main body of the DOS has a Gaussian form and, hence, is symmetric. Asymmetry of the DOS for any arbitrary QG (being, in general, a mixture of AQ and PQ) reflects the particular geometry of a quadrupole and is directly related to the contribution of the axial component. This is not so for DGs, where the inversion of the dipole moment is equivalent to the spatial rotation and the DOS is exactly symmetric.

\subsection{Axial quadrupoles}

DOS can be calculated using \eq{distribution} with the dipolar energy replaced by the quadrupolar one
\begin{equation}\label{Uq-axial}
U_n=u_n\left(3\cos^2\vartheta_n-1\right),\hskip10pt u_n=\frac{eQ}{4\varepsilon r_n^3},
\end{equation}
and
\begin{equation}\label{distribution_Q}
 S(y)=\sum\limits_n\ln\left[\frac{c}{2}\int\limits_{-1}^1 dx \hskip2pt\exp\left[iy u_n(3x^2-1)\right]+1-c\right], \hskip10pt x=\cos\vartheta_n.
\end{equation}
Again, at the saddle point $y_s=i\omega$. Replacing summation over $n$ with integration over $z=\frac{eQ\omega}{4\varepsilon r^3_n}$ we obtain
\begin{eqnarray}\label{S(om)}
S(\omega)\simeq \frac{4\pi}{3}z_a\int\limits_0^{z_a} \frac{dz}{z^2}
\ln\left[cF_a(z)+1-c\right],\hskip10pt z_a=|\omega|U_a=\frac{eQ|\omega|}{4\varepsilon a^3},\\
F_a(z)=\int\limits_0^1 dx \exp\left[-z(3x^2-1)\right].
\end{eqnarray}
For $|U|\rightarrow \infty$ saddle point is located at $|\omega|\rightarrow \infty$. Let us consider the case $\omega\rightarrow\infty$. Then
\begin{equation}\label{F+}
F_a(z)\rightarrow \frac{1}{2}\left(\frac{\pi}{3z}\right)^{1/2}e^z.
\end{equation}
Hence, keeping the major contribution and leading correction only
\begin{equation}\label{S+}
S(\omega)\approx\frac{4\pi}{3}z_a\left(\ln z_a+A_q\right),
\end{equation}
where
\begin{equation}\label{A}
A_q=\int\limits_0^{1} \frac{dz}{z^2}
\ln\left[cF_a(z)+1-c\right]+
\int\limits_1^{\infty} \frac{dz}{z^2}
\ln\left[cF_a(z)e^{-z} +(1-c)e^{-z}\right].
\end{equation}
Saddle point equation is
\begin{equation}\label{Sp+}
-U+\frac{4\pi}{3}\left(\ln z_a+A_q+1\right)U_a=0
\end{equation}
(we see that $U > 0$ for $\omega >0$), and the final asymptotics is
\begin{equation}\label{P+}
P(U)\simeq \left(\frac{3z_a}{8\pi^2 U_a^2}\right)^{1/2}\exp\left(-\frac{4\pi}{3}z_a\right),
\hskip10pt z_a=\exp\left(\frac{3}{4\pi}\frac{U}{U_a}-A_q-1\right),\hskip10pt U\rightarrow \infty.
\end{equation}

Analogous calculation for $\omega\rightarrow -\infty$ provides the asymptotics for $U\rightarrow -\infty$. Here the corresponding limit for the function $F_a(z)$ is
\begin{equation}\label{F-}
F_a(z)\rightarrow -\frac{c}{6z}e^{-2z}, \hskip10pt z=\omega u_n,
\end{equation}
\begin{equation}\label{S-}
S(\omega)\approx \frac{4\pi}{3}z_a\left(2\ln z_a+B_q\right),
\end{equation}
and finally
\begin{equation}\label{P-}
P(U)\simeq \frac{(3z_a)^{1/2}}{4\pi U_a}\exp\left(-\frac{8\pi}{3}z_a\right),
\hskip10pt z_a=\exp\left(-\frac{3}{8\pi}\frac{U}{U_a}-B_q/2-1\right), \hskip10pt U\rightarrow -\infty,
\end{equation}
where
\begin{equation}\label{B}
B_q=\int\limits_0^{1} \frac{dz}{z^2}
\ln\left[c F_a(-z) +1-c\right]+
\int\limits_1^{\infty} \frac{dz}{z^2}
\ln\left[cF_a(-z)e^{-2z}+(1-c)e^{-2z}\right].
\end{equation}
We see that for axial quadrupoles the far tails of DOS are asymmetric in agreement with the general consideration in the previous section.

\subsection{Planar quadrupoles}

For planar quadrupoles
\begin{equation}\label{Uq-planar}
U_n=u_n\left(\cos^2\vartheta_n-\sin^2\vartheta_n\cos^2\varphi_n\right),\hskip10pt u_n=\frac{eQ}{2\varepsilon r_n^3},
\end{equation}
and
\begin{equation}\label{Sn-planar}
S(y)=\sum\limits_n\ln\left[\frac{c}{2\pi}\int\limits_0^1 dx \int\limits_0^{2\pi}d\varphi\hskip2pt\exp\left[iy u_n\left(x^2\left(1+\cos^2\varphi\right)-\cos^2\varphi\right)\right]+1-c\right].
\end{equation}
For $y_s=i\omega$ and $\omega\rightarrow \infty$ the corresponding function ($z=\omega u_n$)
\begin{eqnarray}\label{Sn-planar1}
F_p(z)=\frac{1}{2\pi}\int\limits_0^1 dx \int\limits_0^{2\pi}d\varphi\hskip2pt\exp\left[-z\left(x^2\left(1+\cos^2\varphi\right)-\cos^2\varphi\right)\right]\approx\\
\approx\frac{1}{4\left(z\pi\right)^{1/2}} \int\limits_0^{2\pi}d\varphi\hskip2pt
\frac{\exp\left(z\cos^2\varphi\right)}{\left(1+\cos^2\varphi\right)^{1/2}}\approx
\frac{e^z}{2\sqrt{2}z}.
\end{eqnarray}
For large $\omega$ the asymptotics of $S(\omega)$ is
\begin{equation}\label{Splanar}
S(z_a)\approx\frac{4\pi}{3}z_a\left(\ln z_a+C_q\right),\hskip10pt z_a=\omega U_p=\frac{eQ\omega}{2\varepsilon a^3},
\end{equation}
where
\begin{equation}\label{C}
C_q=\int\limits_0^{1} \frac{dz}{z^2}
\ln \left[cF_p(z) +1-c\right]+
\int\limits_1^{\infty} \frac{dz}{z^2}
\ln\left[cF_p(z)e^{-z}+(1-c)e^{-z}\right]
\end{equation}
and this asymptotics is valid for $\omega\rightarrow \pm\infty$ if we define $z_a=|\omega|U_p$ in full agreement with the general symmetry of the DOS. Hence, the asymptotics of $P(U)$ for planar quadrupoles is described by \eq{P+} where $U$ is replaced by $|U|$, $U_a$ by $U_p$, and $A_q$ by $C_q$. For planar quadrupoles this asymptotics is valid for both positive and negative $U$.

Analogous to the DG model, saddle point approximation is valid for axial and planar quadrupoles if $\left|\frac{d^3 S}{d\omega^3}\right|\left(\frac{d^2 S}{d\omega^2}\right)^{-3/2} \propto z_a^{-1/2} \ll 1$. This inequality is again valid if $|U|/U_{a,p} \gg 1$.

Arbitrary quadrupole is a linear combination of axial and planar quadrupoles. For the corresponding QG model the actual far tail asymptotics is determined by the contribution of the component providing the slowest decay of DOS.

\section{Discussion: implication for charge carrier transport}
Main body of DOS in DG and QG models is Gaussian and, thus, symmetric. This means that the effect of the energetic disorder on the transport of electrons and holes is exactly the same if the transport is determined by the main body of DOS. Analytic results and computer simulation show that the tails of DOS at $|U|/\sigma\simeq 4-5$ are still Gaussian \cite{Novikov:877e,Dieckmann:8136}. Far tails in the QG model becomes asymmetric and provide inequivalent environment for carriers of the opposite signs. Unfortunately, experimental observation of the quasi-equilibrium charge transport for low temperature $\sigma/kT\simeq 7-8$ is extremely difficult.

Still, we could provide some estimations for the low field mobility temperature dependence in the DG model for low temperature. Using the approach described in the Introduction and estimating the effective activation energy using the position of the maximum of the occupied DOS, we obtain for the leading asymptotics
\begin{equation}\label{lowT-DG}
\ln\mu\propto 4\pi \frac{U_d}{kT}-3\pi A_d\left(\frac{U_d}{kT}\right)^{3/2}.
\end{equation}
This estimation is valid when $U_d/kT \gg 1$ and the second term in \eq{lowT-DG} is dominating. For the axial QG model the corresponding relation is (for one particular side of DOS)
\begin{equation}\label{lowT-aQG}
\ln\mu\propto -\frac{4\pi}{3} \frac{U_a}{kT}\left[A_q+1+\ln\left(\frac{U_a}{ kT}\right)\right]
\end{equation}
with the trivial modification for the other side of the axial QG DOS or for the case of planar QG. In all cases the temperature dependence becomes weaker in comparison with the Gaussian DOS due to the more faster decay of the DOS tail.

In fact, weakening of the dependence $\mu(T)$ for the time-of-flight experiments at low temperature has been observed previously \cite{Borsenberger:12145,Borsenberger:4289}, but the experimental photocurrent transients are so dispersive that this effect is better attributed to the non-equilibrium transport.  Borsenberger et al. \cite{Borsenberger:12145} showed that in the popular Gaussian Disorder Model transition from the non-dispersive to dispersive transport leads to the change of the low field $\mu(T)$ dependence from $\ln\mu\approx -(2\sigma/3kT)^2$ to $\ln\mu\approx -(\sigma/2kT)^2$, e.g. the slope of the $\ln\mu$ vs $1/T^2$ dependence becomes smaller by the factor $\approx 2$. In some situations in the low temperature region the decrease of the slope becomes much greater (see Fig. 11 in \cite{Mozer:35214}), indicating a possibility of the deviation of the shape of the DOS from the Gaussian one.

Measurement of the  stationary space charge limited current \cite{Mensfoort:85208} or technique of charge carrier extraction by linearly increasing voltage (CELIV) \cite{Mozer:35214} makes it possible to study the quasi-equilibrium transport even for very low temperature $\sigma/kT\simeq 10$. Unfortunately, analysis of the transport data is not so straightforward as in the case of time-of-flight data.

Another important problem is the modification of the mobility field dependence for the non-Gaussian DOS, it will be considered in a separate paper.

\section{Conclusion}

We consider the shape of the far tails of the DOS in the models of amorphous organic materials, suitable for the description of charge carrier transport in polar and nonpolar materials (DG or QG model, correspondingly). It was found that for both models the shape of the tails becomes non-Gaussian, but there is a principal difference between two models: in the DG model the DOS is exactly symmetric around $U=0$ while in the QG model the DOS is generally asymmetric leading to the inequivalence of the energetic landscape for transport of electrons and holes.

Deviation of the DOS shape from the Gaussian one naturally leads to the alteration of the mobility temperature dependence for low temperature: the dependence $\mu(T)$ deviates from the genuine Gaussian dependence $\ln\mu\propto -(\sigma/kT)^2$ and becomes weaker (see \eq{lowT-DG} and (\ref{lowT-aQG})). Observation of such dependence could be a serious argument in favor of the non-Gaussian DOS in the tail region. Unfortunately, any feasible transport experiment should be carried out at the extremely low temperature where time-of-flight transport becomes highly dispersive even for rather thick transport layers. For the quasi-stationary methods such as CELIV or space charge limited current-voltage measurements an extraction of the carrier mobility from the experimental data is not so straightforward.

\section{Acknowledgement}
Financial support from the RFBR grant 15-02-03082 is gratefully acknowledged.


\end{document}